\documentstyle[11pt,epsf,psfig,amsmath,axodraw]{article}
\def\Year{\expandafter\eatPrefix\the\year}
\newcount\hours \newcount\minutes
\def\monthname{\ifcase\month\or
January\or February\or March\or April\or May\or June\or July\or
August\or September\or October\or November\or December\fi}
\def\shortmonthname{\ifcase\month\or
Jan\or Feb\or Mar\or Apr\or May\or Jun\or Jul\or
Aug\or Sep\or Oct\or Nov\or Dec\fi}

\def\TimeStamp{\hours\the\time\divide\hours by60%
\minutes -\the\time\divide\minutes by60\multiply\minutes by60%
\advance\minutes by\the\time%
${\rm \shortmonthname}\cdot   \if\day<10{}0\fi\the\day\cdot   \the\year
\qquad\the\hours:\if\minutes<10{}0\fi\the\minutes$}

\newskip\humongous \humongous=0pt plus 1000pt minus 100pt
\def\caja{\mathsurround=0pt}
\def\eqalign#1{\,\vcenter{\openup1\jot \caja
       \ialign{\strut \hfil$\displaystyle{##}$&$
        \displaystyle{{}##}$\hfil\crcr#1\crcr}}\,}
\newif\ifdtup

\newcounter{eqnumber}[section]
\renewcommand{\theeqnumber}{\thesection.\arabic{eqnumber}}
\def\equn{\refstepcounter{eqnumber}
\eqno({\rm \theeqnumber})
}

\def\npb#1#2#3{{\rm Nucl.\ Phys. } {B\ \bf  #1}, #3 (#2)}
\def\plb#1#2#3{{\rm Phys.\ Lett. } {B\ \bf  #1}, #3 (#2)}
\def\prd#1#2#3{{\rm Phys.\ Rev. } {D\ \bf  #1}, #3 (#2)}

\def\cqg#1#2#3{{\rm Class.\ and Quant.\ Grav.} {\bf  #1}, #3 (#2)}

\def\hepth#1{[hep-th/#1]}
\def\hepph#1{[hep-ph/#1]}

\def\tr{\mathop{\rm tr}\nolimits}

\def\Is#1#2{I
^{{#1}}_{4:#2}}
\def\Ione{\Is{\rm 1m}}
\def\Ieasy{\Is{{\rm 2m}\,e}}
\def\Ihard{\Is{{\rm 2m}\,h}}

\def\Fn{n}
\def\Fs#1#2{F^{{#1}}_{\Fn:#2}}
\def\Fone{\Fs{\rm 1m}}
\def\Feasy{\Fs{{\rm 2m}\,e}}
\def\Fhard{\Fs{{\rm 2m}\,h}}
\def\Fthree{\Fs{\rm 3m}}
\def\Ffour{\Fs{\rm 4m}}

\newbox\charbox
\newbox\slabox
\def\s#1{{      % Feynman slash
        \setbox\charbox=\hbox{$#1$}
        \setbox\slabox=\hbox{$/$}
        \dimen\charbox=\ht\slabox
        \advance\dimen\charbox by -\dp\slabox
        \advance\dimen\charbox by -\ht\charbox
        \advance\dimen\charbox by \dp\charbox
        \divide\dimen\charbox by 2
        \raise-\dimen\charbox\hbox to \wd\charbox{\hss/\hss}
        \llap{$#1$}
}}

\def\spa#1.#2{\left\langle#1\,#2\right\rangle}
\def\spb#1.#2{\left[#1\,#2\right]}
\def\lor#1.#2{\left(#1\,#2\right)}

\catcode`@=11  % Make @ letter.

\def\Slash#1{\hskip 0.05 cm \slash\hskip -0.22 cm #1}

\def\Tr{\, {\rm Tr}}

\def\eps{\epsilon}

\def\e{\epsilon}

\def\la{\langle}
\def\ra{\rangle}

\def\lsl{\not{\hbox{\kern-2.3pt $\ell$}}}
\def\ksl{\not{\hbox{\kern-2.3pt $k$}}}

\def\rg{r_{\Gamma}}

\def\spa#1.#2{\left\langle#1\,#2\right\rangle}
\def\spb#1.#2{\left[#1\,#2\right]}
\def\lor#1.#2{\left(#1\,#2\right)}

\def\sand#1.#2.#3{%
  \left\langle\smash{#1}{\vphantom1}\right|{#2}%
  \left|\smash{#3}{\vphantom1}\right\rangle}
\def\sandp#1.#2.#3{%
  \left\langle\smash{#1}{\vphantom1}^{-}\right|{#2}%
  \left|\smash{#3}{\vphantom1}^{+}\right\rangle}
\def\sandpp#1.#2.#3{%
  \left\langle\smash{#1}{\vphantom1}^{+}\right|{#2}%
  \left|\smash{#3}{\vphantom1}^{+}\right\rangle}
\def\sandmm#1.#2.#3{%
  \left\langle\smash{#1}{\vphantom1}^{-}\right|{#2}%
  \left|\smash{#3}{\vphantom1}^{-}\right\rangle}
\def\sandpm#1.#2.#3{%
  \left\langle\smash{#1}{\vphantom1}^{+}\right|{#2}%
  \left|\smash{#3}{\vphantom1}^{-}\right\rangle}
\def\sandmp#1.#2.#3{%
  \left\langle\smash{#1}{\vphantom1}^{-}\right|{#2}%
  \left|\smash{#3}{\vphantom1}^{+}\right\rangle}

\def\Atree{A^{\rm tree}}
\def\Atreemhv{A^{\rm tree\ MHV}}
\def\A#1{{\cal A}_{#1}}

\def\dlips{dLIPS}

\def\tr{\mathop{\hbox{\rm tr}}\nolimits}

\def\L{\left(}\def\R{\right)}

\def\LB{\left[}\def\RB{\right]}

\def\tn#1#2{t^{[#1]}_{#2}}
\def\L{\left(}\def\R{\right)}
\def\dlips{d{\rm LIPS}}

\def\BRR#1#2{\la#1|\Slash{P}|#2\ra}
\def\tree{{\rm tree}}
\def\treemhv{{\rm tree\ MHV}}

\def\Gr{{\rm Gr}}

\begin{document}

\begin{titlepage}

\begin{flushright}
%\today
%\TimeStamp
%\\
hep-th/0412023 
\\
SWAT-04-420 \\
\end{flushright}

\vskip 2.cm

\begin{center}
\begin{Large}
{\bf Twistor Space Structure of the Box Coefficients of $N=1$ One-loop Amplitudes}

\vskip 2.cm

\end{Large}

\vskip 2.cm

{\large
Steven J. Bidder %${}^{1,\dagger}$
,
N.\ E.\ J.\ Bjerrum-Bohr %${}^{1,\dagger}$
,
David C. Dunbar %$^{1,\dagger}$
and Warren B. Perkins
} %$^{}$}

\vskip 0.5cm

{\it %${}^1$
Department of Physics, \\
University
    of Wales Swansea,
\\ Swansea, SA2 8PP, UK }

\vskip .3cm

\begin{abstract}
We examine the coefficients of the box functions in $N=1$ supersymmetric one-loop amplitudes.
We present the box coefficients for all six point $N=1$ amplitudes and certain all $n$ example 
coefficients. 
We find for ``next-to MHV'' amplitudes that these box coefficients have coplanar support in twistor space.

\end{abstract}

\end{center}

\vfill
\noindent\hrule width 3.6in\hfil\break
%
%${}^{\dagger}$ 
%Research supported by the PPARC.\hfil\break
%
\{pysb, n.e.j.bjerrum-bohr, d.c.dunbar, w.perkins\}@swan.ac.uk

\end{titlepage}

\section{Introduction}

Recently a ``weak-weak'' duality has been proposed
between $N=4$ supersymmetric gauge theory and topological string
theory~\cite{Witten:2003nn}. This relationship becomes manifest by
transforming amplitudes into twistor space where they are supported on
simple curves.  A consequence of this picture is that tree amplitudes,
when expressed as functions of spinor variables $k_{a\dot a} =\lambda_a
\tilde\lambda_{\dot a}$, are annihilated by various differential
operators corresponding to the localization of points to lines and
planes in twistor space.  In particular the operator corresponding to
collinearity of points $i,j,k$ in twistor space is
$$
[F_{ijk} , \eta ] =
 \spa{i}.j \left[{ \partial\over \partial\tilde\lambda_k},\eta\right]
+\spa{j}.k \left[{ \partial\over
\partial\tilde\lambda_i},\eta\right] +\spa{k}.i \left[{
\partial\over
\partial\tilde\lambda_j},\eta\right]
\equn
$$
and similarly annihilation by 
$$
K_{ijkl} =
\spa{i}.{j} \epsilon^{\dot a \dot b}
{ \partial \over \partial \tilde \lambda_k^{\dot a}  }
{ \partial \over \partial \tilde \lambda_l^{\dot b}  }
+{\rm perms} \equn
$$
indicates {\it co-planarity} of points $i,j,k$ and $l$, {\it i.e.,} these four
points lie on a plane in twistor space.

As an example, the collinear operator
annihilates  the
``maximally helicity violating'' (MHV) $n$-gluon tree amplitudes,
$$
[F_{ijk} , \eta ] A_n^{\rm tree\; MHV}(1,2,\cdots n) = 0,
\equn 
$$
indicating that such amplitudes only have non-zero support on a line in twistor
space.  These MHV colour-ordered amplitudes, where exactly two of the
gluons have negative helicity, have a remarkably simple form
(see equation~(\ref{ParkeTaylor})), conjectured by Parke and
Taylor~\cite{ParkeTaylor} and proved by Berends and
Giele~\cite{BerendsGiele}. 
Using ``cut-constructibility'' and collinear limits, the one-loop MHV
amplitudes have also been constructed for $N=4$~\cite{BDDKa} and 
$N=1$ supersymmetric theories~\cite{BDDKb}. 
``Cut-constructible'' implies that the entire amplitude can be
reconstructed from a knowledge of its four-dimensional cuts~\cite{BDDKa,BDDKb,BernMorgan}.
The MHV tree amplitudes appear to play a key role in gauge
theories.  Cachazo, Svr\v{c}ek and Witten have conjectured that
Yang-Mills amplitudes could be calculated using off-shell MHV
vertices~\cite{Cachazo:2004kj}. 
This construction can be extended to other particle types~\cite{CSW:matter}
and has already had multiple applications~\cite{Tree}.

An understanding of the twistor structure of loop amplitudes has
proven more difficult. However, Brandhuber et.
al.~\cite{Brandhuber:2004yw} demonstrated by computation of the $N=4$
MHV $n$-point amplitudes how the CSW construction can be extended to
one-loop amplitudes. Furthermore, the twistor space structure has been shown to
manifest itself in the coefficients of the integral functions defining
one-loop amplitudes. For $N=4$ one-loop amplitudes where only box
integral functions appear, the coefficients of these box functions
satisfy co-planarity and
collinearity conditions in twistor space~\cite{BeDeDiKo,Britto:2004tx}.  
These twistor space
conditions have been shown to be
useful in determining the coefficients of  $N=4$ one-loop
amplitudes~\cite{Cachazo:2004by,Cachazo:2004dr,Britto:2004nj,Bena:2004xu}.

For $N=1$ supersymmetric one-loop amplitudes much less is known.  In
refs.~\cite{Quigley:2004pw,Bedford:2004py} it was shown that the CSW
constructions~\cite{Cachazo:2004kj,CSW:matter} can  be employed in computing one-loop $N=1$ amplitudes by
reproducing the MHV $n$-point amplitude and in ref.~\cite{BBDD} it was
shown how the holomorphic anomaly applies to $N=1$ amplitudes.  $N=1$
amplitudes have a more complicated structure than $N=4$ amplitudes,
containing box, triangle and bubble integral functions. In this
letter, we present the box coefficients of all six point
one-loop $N=1$ amplitudes and  some specific box
coefficients in $n$-point amplitudes. 
We find that, as in the case of $N=4$, for amplitudes with three
negative helicities (``next to MHV'') the box coefficients have
coplanar support in twistor space, while for $q>3$ negative helicities
this simple behaviour is no longer true.

\section{Organization of One-Loop Gauge Theory Amplitudes}

Tree-level amplitudes for $U(N_c)$ or $SU(N_c)$ gauge theories with
$n$ external gluons can be decomposed into colour-ordered partial
amplitudes multiplied by an associated
colour-trace~\cite{ChanPaton,TreeColour,ManganoReview}.  Summing over
all non-cyclic permutations reconstructs the full amplitude
$\A{n}^\tree$ from the partial amplitudes $A_n^\tree(\sigma)$,
$$
\A{n}^\tree(\{k_i,\lambda_i,a_i\}) =
g^{n-2} \sum_{\sigma\in S_n/Z_n} \Tr(T^{a_{\sigma(1)}}
\cdots T^{a_{\sigma(n)}})
\ A_n^\tree(k_{\sigma(1)}^{\lambda_{\sigma(1)}},\ldots,
            k_{\sigma(n)}^{\lambda_{\sigma(n)}})\ ,
\equn\label{TreeAmplitudeDecomposition}
$$
where $k_i$, $\lambda_i$, and $a_i$ are respectively the momentum,
helicity ($\pm$) and colour-index of the $i$-th external
gluon, $g$ is the coupling constant and $S_n/Z_n$ is the set of
non-cyclic permutations of $\{1,\ldots, n\}$.
The $U(N_c)$ ($SU(N_c)$) generators $T^a$ are the set of hermitian
(traceless hermitian) $N_c\times N_c$ matrices,
normalized such that $\Tr\L T^a T^b\R = \delta^{ab}$.
The colour decomposition~(\ref{TreeAmplitudeDecomposition})
can be derived in conventional field theory simply by using
$
f^{abc} = -i \Tr\L \LB T^a, T^b\RB T^c\R/\sqrt2,
$
where the $T^a$ may be either $SU(N_c)$ matrices or $U(N_c)$ matrices.

In a supersymmetric theory amplitudes with all helicities identical,
or with all but one helicity identical, vanish due to supersymmetric Ward
identities~\cite{SWI}.
Tree-level gluon amplitudes in super-Yang-Mills and in
purely gluonic Yang-Mills are identical
(fermions do not appear at this order), so that
$$
A_n^\tree(1^{\pm},2^{+}, \ldots,n^+) = 0.\equn
$$
The non-vanishing Parke-Taylor formula~\cite{ParkeTaylor} for
the MHV partial amplitudes is,
$$
\eqalign{
  A_{jk}^\treemhv(1,2,\ldots,n)\
\equiv\
  A^\tree_n(1^+,\ldots,j^-,\ldots,k^-,
                \ldots,n^+),
&=\ i\, { {\spa{j}.{k}}^4 \over \spa1.2\spa2.3\cdots\spa{n}.1 }\ ,
  \cr}\equn
\label{ParkeTaylor}
$$
for a partial amplitude where $j$ and $k$ are the only legs with
negative helicity. In our convention all legs are outgoing.
The result~(\ref{ParkeTaylor}) is written in terms of spinor inner-products,
$\spa{j}.{l} \equiv \langle j^- | l^+ \rangle $,
$\spb{j}.{l} \equiv \langle j^+ | l^- \rangle $,
where $| i^{\pm}\ra $ is a massless Weyl spinor with momentum $k_i$ and
chirality $\pm$~\cite{SpinorHelicity,ManganoReview}. In terms of the variables
$\lambda^i_{m},\tilde\lambda^i_{\dot m}$,
$$
\spa{i}.j = \eps^{mn}\lambda^i_{m}\lambda^j_{n}
\;  ,  
\;\; \;\;
\spb{i}.j = \eps^{\dot m\dot n}\tilde\lambda^i_{\dot m}\tilde\lambda^j_{\dot n}.
\equn
$$
The spinor products are related to the momentum invariants by 
$\spa{i}.j\spb{j}.i=2k_i \cdot k_j\equiv s_{ij}$ with $(\spa{i}.j)^*=\spb{j}.i$.
The tree amplitudes contain many residual features of higher symmetries~\cite{Nair}.
For one-loop amplitudes, one may perform a similar colour decomposition
to the tree-level decomposition~(\ref{TreeAmplitudeDecomposition})~\cite{Colour}.
In this case there are  two traces
over colour matrices
and one must also sum over the different spins, $J$, of the internal
particles circulating in the loop.
When all particles transform as colour adjoints,
the result takes the form,
$$
{\cal A}_n\L \{k_i,\lambda_i,a_i\}\R =
g^n  \sum_{J}\,\sum_{c=1}^{\lfloor{n/2}\rfloor+1}
      \sum_{\sigma \in S_n/S_{n;c}}
     \Gr_{n;c}\L \sigma \R\,A_{n;c}^{[J]}(\sigma),
\label{ColourDecomposition}\equn
$$
where ${\lfloor{x}\rfloor}$ is the largest integer less than or equal to $x$.
The leading colour-structure factor,
$$
\Gr_{n;1}(1) = N_c\ \Tr\L T^{a_1}\cdots T^{a_n}\R \, ,\equn
$$
is just $N_c$ times the tree colour factor, and the subleading colour
structures ($c>1)$ are given by
$$
\Gr_{n;c}(1) = \Tr\L T^{a_1}\cdots T^{a_{c-1}}\R\,
\Tr\L T^{a_c}\cdots T^{a_n}\R \, .\equn
$$
$S_n$ is the set of all permutations of $n$ objects,
and $S_{n;c}$ is the subset leaving $\Gr_{n;c}$ invariant.
Once again it is convenient to use $U(N_c)$ matrices; the extra $U(1)$
decouples~\cite{Colour}.
(For internal particles in the fundamental ($N_c+\bar{N_c}$) representation,
only the single-trace colour structure ($c=1$) would be present,
and the corresponding colour factor would be smaller by a factor of $N_c$.
In this case the $U(1)$ gauge boson will {\it not} decouple from
the partial amplitude, so one should only sum over $SU(N_c)$ indices
when colour-summing the cross-section.)

For one-loop amplitudes, the subleading in colour amplitudes 
$A_{n;c}   \;c > 1 $
may be obtained by summations of permutations
of the leading in colour amplitude~\cite{BDDKa},
$$
 A_{n;c}(1,2,\ldots,c-1;c,c+1,\ldots,n)\ =\
 (-1)^{c-1} \sum_{\sigma\in COP\{\alpha\}\{\beta\}} A_{n;1}(\sigma),
\equn
$$
where $\alpha_i \in \{\alpha\} \equiv \{c-1,c-2,\ldots,2,1\}$,
$\beta_i \in \{\beta\} \equiv \{c,c+1,\ldots,n-1,n\}$,
and $COP\{\alpha\}\{\beta\}$ is the set of all
permutations of $\{1,2,\ldots,n\}$ with $n$ held fixed
that preserve the cyclic
ordering of the $\alpha_i$ within $\{\alpha\}$ and of the $\beta_i$
within $\{\beta\}$, while allowing for all possible relative orderings
of the $\alpha_i$ with respect to the $\beta_i$.
Hence, we need only focus on the leading in colour amplitude and use this
relationship to generate the full amplitude if required.

For $N=1$ super Yang-Mills with external gluons there are two
possible supermultiplets contributing to the one-loop amplitude:
the vector and the chiral matter multiplets, 
which can be decomposed into single particle contributions,
$$
\eqalign{ A_{n}^{N=1\ {\rm vector}}\ \equiv\ A_{n}^{[1]}\
+A_{n}^{[1/2]} \,, \cr A_{n}^{N=1\ {\rm chiral}}\ \equiv\
A_{n}^{[1/2]}\ +A_{n}^{[0]} \, . \cr}\equn
$$
For spin-$0$ we always consider a complex scalar. 
Throughout we assume the use of a supersymmetry preserving 
regulator~\cite{Siegel,StringBased,KST}.
For $N=4$
super Yang-Mills theory there is a single multiplet
whose contribution is given by
$$
A_{n}^{N=4}\ \equiv\
A_{n}^{[1]} + 4A_{n}^{[1/2]}+3 A_{n}^{[0]}\,.\equn
$$
The contributions from the three supersymmetric multiplets
are not independent but satisfy
$$
A_{n}^{N=1\ {\rm vector}}\ \equiv\ A_{n}^{N=4} -
3A_{n}^{N=1\ {\rm chiral}} \,.\equn
$$
Thus, provided the $N=4$ amplitude is known, one need only
calculate one of the two possibilities for $N=1$.  The $N=4$
six-point amplitudes are known~\cite{BDDKa,BDDKb} and their
twistor space structure has been examined. In this letter
we focus on the
$A_{6}^{N=1\ {\rm chiral}}$ amplitudes.

\section{Basis of Functions}

In general, one-loop amplitudes can
be decomposed in terms of a set of basis functions, $I_i$, with 
coefficients, $c_i$, that are rational in terms of spinor products,
$$
A = \sum_{i}  c_i  I_i \,.
\label{generalform2}\equn
$$
In a Feynman diagram calculation the coefficients may, in
principle, be obtained from a Passerino-Veltman reduction~\cite{PassVelt}.  
For
supersymmetric amplitudes, the set is  restricted  due to
cancellations within the loop-momentum integrals.  For $N=1$
amplitudes the set can be taken to contain scalar boxes, $I_4$, scalar
triangles, $I_3$,  and scalar bubbles, $I_2$.
In this letter we  focus on the behaviour of the box
functions. 
In general, we can organize the box functions according to the
number of legs with non-null input momenta and the relative
labeling of legs.  Specifically we have,
$$
  I_{4:i}^{1{\rm m}}
\hskip 0.5truecm
 I_{4:r;i}^{2{\rm m}e}
\hskip 0.5truecm
  I_{4:r;i}^{2{\rm m}h}
\hskip 0.5truecm
  I_{4:r,r',i}^{3{\rm m}}
\hskip 0.5truecm
I_{4: r, r', r'', i}^{4{\rm m}} \equn
$$
with the labeling as indicated,

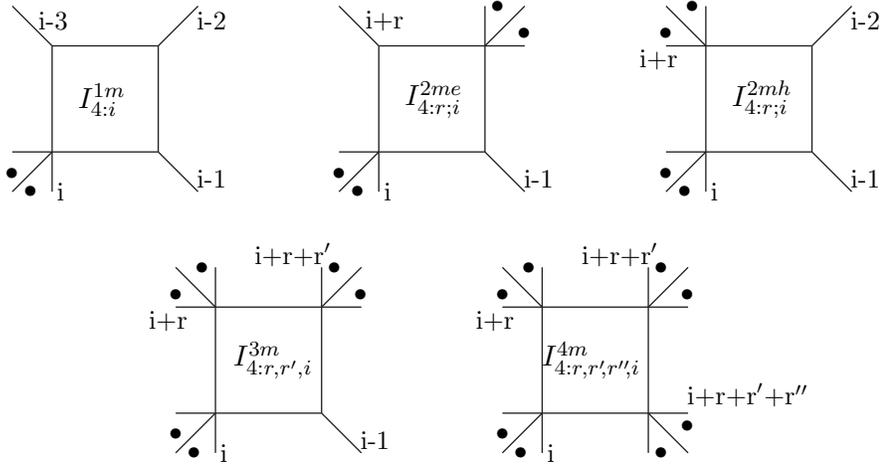
\begin{center}

\begin{picture}(120,95)(0,0)
%Picture 1 1 mass
% Lines of Box
\Line(30,20)(70,20)
\Line(30,60)(70,60)
\Line(30,20)(30,60)
\Line(70,60)(70,20)

%Lines of corner lines
\Line(30,20)(15,5)
\Line(30,60)(15,75)
\Line(70,20)(85,5)
\Line(70,60)(85,75)

%Massive Lines
\Line(30,20)(30,5)
\Line(30,20)(15,20)
\Text(22,5)[c]{\small$\bullet $}
\Text(15,12)[c]{\small $\bullet $}

\Text(32,5)[l]{\small $\hbox{\rm i}$}
\Text(85,10)[l]{\small $\hbox{\rm i-1}$}
\Text(85,70)[l]{\small $\hbox{\rm i-2}$}
\Text(25,70)[l]{\small $\hbox{\rm i-3}$}

\Text(40,40)[l]{$I^{1m}_{4:i}$}
\end{picture}
\begin{picture}(120,95)(0,0)
%Picture 2: 2 mass easy
% Lines of Box
\Line(30,20)(70,20)
\Line(30,60)(70,60)
\Line(30,20)(30,60)
\Line(70,60)(70,20)

%Lines of corner lines
\Line(30,20)(15,5)
\Line(30,60)(15,75)
\Line(70,20)(85,5)
\Line(70,60)(85,75)

%Massive Lines
\Line(30,20)(30,5)
\Line(30,20)(15,20)
\Text(22,5)[c]{\small$\bullet$}
\Text(15,12)[c]{\small $\bullet$}

%Massive Lines
\Line(70,60)(70,75)
\Line(70,60)(85,60)
\Text(75,75)[c]{\small$\bullet$}
\Text(85,65)[c]{\small $\bullet$}

\Text(32,5)[l]{\small $\hbox{\rm i}$}
\Text(85,10)[l]{\small $\hbox{\rm i-1}$}
\Text(25,70)[l]{\small $\hbox{\rm i+r}$}

\Text(40,40)[l]{$I^{2me}_{4:r;i}$}
\end{picture}
\begin{picture}(120,95)(0,0)
% Picture 3 2 mass hard
% Lines of Box
\Line(30,20)(70,20)
\Line(30,60)(70,60)
\Line(30,20)(30,60)
\Line(70,60)(70,20)

%Lines of corner lines
\Line(30,20)(15,5)
\Line(30,60)(15,75)
\Line(70,20)(85,5)
\Line(70,60)(85,75)

%Massive Lines
\Line(30,20)(30,5)
\Line(30,20)(15,20)
\Text(22,5)[c]{\small$\bullet$}
\Text(15,12)[c]{\small $\bullet$}

%Massive Lines
\Line(30,60)(30,75)
\Line(30,60)(15,60)
\Text(15,65)[c]{\small$\bullet$}
\Text(25,75)[c]{\small $\bullet$}

\Text(32,5)[l]{\small $\hbox{\rm i}$}
\Text(85,10)[l]{\small $\hbox{\rm i-1}$}
\Text(85,70)[l]{\small $\hbox{\rm i-2}$}
\Text(5,55)[l]{\small $\hbox{\rm i+r}$}

\Text(40,40)[l]{$I^{2mh}_{4:r;i}$}
\end{picture}

\begin{picture}(120,95)(0,0)
% Picture 4 3 mass hard
% Lines of Box
\Line(30,20)(70,20)
\Line(30,60)(70,60)
\Line(30,20)(30,60)
\Line(70,60)(70,20)

%Lines of corner lines
\Line(30,20)(15,5)
\Line(30,60)(15,75)
\Line(70,20)(85,5)
\Line(70,60)(85,75)

%Massive Lines
\Line(30,20)(30,5)
\Line(30,20)(15,20)
\Text(22,5)[c]{\small $\bullet$}
\Text(15,12)[c]{\small $\bullet$}

%Massive Lines
\Line(30,60)(30,75)
\Line(30,60)(15,60)
\Text(15,65)[c]{\small $\bullet$}
\Text(25,75)[c]{\small $\bullet$}

%Massive Lines
\Line(70,60)(70,75)
\Line(70,60)(85,60)
\Text(75,75)[c]{\small $\bullet$}
\Text(85,65)[c]{\small $\bullet$}

\Text(32,5)[l]{\small $\hbox{\rm i}$}
\Text(85,10)[l]{\small $\hbox{\rm i-1}$}
\Text(45,80)[l]{\small $\hbox{\rm i+r+r}'$}
\Text(5,55)[l]{\small $\hbox{\rm i+r}$}

\Text(37,40)[l]{$I^{3m}_{4:r,r',i}$}
\end{picture}
\begin{picture}(120,95)(0,0)
% Picture 4 4 mass 
% Lines of Box
\Line(30,20)(70,20)
\Line(30,60)(70,60)
\Line(30,20)(30,60)
\Line(70,60)(70,20)

%Lines of corner lines
\Line(30,20)(15,5)
\Line(30,60)(15,75)
\Line(70,20)(85,5)
\Line(70,60)(85,75)

%Massive Lines
\Line(30,20)(30,5)
\Line(30,20)(15,20)
\Text(22,5)[c]{\small $\bullet$}
\Text(15,12)[c]{\small $\bullet$}

%Massive Lines
\Line(30,60)(30,75)
\Line(30,60)(15,60)
\Text(15,65)[c]{\small $\bullet$}
\Text(25,75)[c]{\small $\bullet$}

%Massive Lines
\Line(70,60)(70,75)
\Line(70,60)(85,60)
\Text(75,75)[c]{\small $\bullet$}
\Text(85,65)[c]{\small $\bullet$}

%Massive Lines
\Line(70,20)(70,5)
\Line(70,20)(85,20)
\Text(75,5)[c]{\small $\bullet$}
\Text(85,15)[c]{\small $\bullet$}

\Text(32,5)[l]{\small $\hbox{\rm i}$}
\Text(85,27)[l]{\small $\hbox{\rm i+r+r}'\hbox{\rm +r}''$}
\Text(45,80)[l]{\small $\hbox{\rm i+r+r}'$}
\Text(5,55)[l]{\small $\hbox{\rm i+r}$}

\Text(30,40)[l]{$I^{4m}_{4:r,r'\hskip -2pt ,r''\hskip -2pt,i}$}
\end{picture}

\end{center}

There is a choice as to
which basis of functions to use, particularly with the bubble
and triangle functions. For the boxes there is rather less freedom.
Nevertheless we can consider three choices of basis, 
each of which has advantages in certain circumstances:

$\bullet$ $D=4$ scalar box integrals, 

$\bullet$ $D=6$ scalar box integrals,

$\bullet$ $D=4$ scalar box  $F$-functions

\noindent
The $D=4$ scalar box integrals are the natural choice, but 
$D=6$ scalar box integrals have 
several practical
advantages. Firstly they are IR finite, which makes determining 
their collinear limits particularly
simple.  
Secondly, for the $N=1$ chiral multiplet, the amplitude has a leading
$\eps^{-1}$ singularity in dimensional
regularization~\cite{Kunszt:1994mc}. 
Scalar triangles have $1/\eps^2$
and $\ln(s)/\eps$ singularities.  As the $D=6$ boxes are IR finite,
there can be no cancellation of the
$\eps^{-2}$ and $\ln(s)/\eps$ terms
between them and the triangles.
This implies the absence of the
scalar triangle integrals in these amplitudes. The relationship
between the $D=4$ boxes and $D=6$ boxes involves an overall factor and
triangles functions, specifically, using the notation of
ref.~\cite{BDKintegrals},
$$
I_4^{D=4}  = { 1 \over 2 N_4 }
\Biggl[
\sum_i \alpha_i \gamma_i I_3^{(i)}  +(-1+2\eps)\hat \Delta_4  I_4^{D=6}
\Biggr].\equn
$$
 The $\hat \Delta_4$ are
rational functions of the momentum invariants,
$$
\eqalign{
{\hat\Delta_{4:i}^{\rm 1m } \over 2N_4 } =&
-2\left( 
{ t_{i-3}^{[2]}+ t_{i-2}^{[2]}- t_{i}^{[n-3]} \over
 t_{i-3}^{[2]}t_{i-2}^{[2]}  } \right)
=2{(k_{i-1}+k_{i-3})^2\over
 (k_{i-3}+k_{i-2})^2(k_{i-2}+k_{i-1})^2
}
\cr
{\hat\Delta_{4:r;i}^{\rm 2mh } \over 2N_4 } =&
-2\left( 
{ 
( t_{i-1}^{[r+1]}- t_i^{[r]})
( t_{i-1}^{[r+1]}- t_{i+r}^{[n-r-2]})
+t_{i-1}^{[r+1]} t_{i-2}^{[2]} )  \over
 t_{i-2}^{[2]} (t_{i-1}^{[r+1]})^2  } \right)
= 
{ \tr ( \Slash{k}_{i-1} \Slash{P}_{i-1\ldots i+r-1} \Slash{k}_{i-2}  \Slash{P}_{i-1\ldots i+r-1} )
\over 
( k_{i-2}+k_{i-1} )^2  (P^2_{i-1\ldots i+r-1} )^2  }
\cr
{\hat\Delta_{4:r;i}^{\rm 2me } \over 2N_4 } =&
-2 \left( {\tn{r+1}{i-1}+\tn{r+1}{i} -\tn{r}{i}-\tn{n-r-2}{i+r+1}\over 
 \tn{r+1}{i-1}\tn{r+1}{i} -\tn{r}{i}\tn{n-r-2}{i+r+1} }\right) 
\cr}\equn
$$
where $t_{a}^{[p]}\equiv (k_a+k_{a+1}+\cdots +k_{a+p-1})^2=P_{a\ldots a+p-1}^2$ and 
$P_{i\ldots j}=k_i+k_{i+1}\ldots +k_{j}$.

 The four
dimensional boxes have dimension $-2$. It is convenient to
define dimension zero $F$-functions by removing the momentum
prefactors of the $D=4$ scalar boxes~\cite{BDDKb},
$$
I_4^{D=4} = { 1 \over K} F_4\ . \equn 
$$
For the $N=4$  amplitudes, it is the coefficients of these $F$-functions which 
the collinearity and co-planarity operators 
annihilate~\cite{Cachazo:2004by,Cachazo:2004dr,BeDeDiKo}.
Explicitly,
$$
\eqalign{
  I_{4:i}^{1{\rm m}} =\ -2 \rg {\Fone{i} \over \tn{2}{i-3} \tn{2}{i-2} }
        \,  \hskip 1 cm
 I_{4:r;i}^{2{\rm m}e}
=\ -2 \rg {\Feasy{r;i}
      \over \tn{r+1}{i-1}\tn{r+1}{i} -\tn{r}{i}\tn{n-r-2}{i+r+1} }\,,
        \,  \hskip 1 cm
  I_{4:r;i}^{2{\rm m}h}
=\ -2 \rg {\Fhard{r;i} \over \tn{2}{i-2} \tn{r+1}{i-1} } \,,
\cr
  I_{4:r,r',i}^{3{\rm m}}
=\ -2 \rg {\Fthree{r,r';i}
     \over \tn{r+1}{i-1} \tn{r+r'}i -\tn{r}{i} \tn{n-r-r'-1}{i+r+r'} }\,,
        \,  \hskip 1 cm
I_{4: r, r', r'', i}^{4{\rm m}}  =
-2 {\Ffour{r, r', r'';i}\over t_i^{[r+ r']}\; t_{i+r}^{[r'+r'']}\;\rho}\, .
\cr}\equn
$$
For  the box functions
it is easy to switch between bases since
$$
A|_{{\rm boxes}} = \sum_i c_i^{D=4}  I^{D=4}_i
=\sum_i c_i^{D=6}  I^{D=6}_i
=\sum_i c_i^{F}  F_i, \equn
$$
thus the coefficients must satisfy
$$
 c_i^{D=4} =   { c_i^{D=6} \over ( - \hat \Delta_4/2 N_4) }
= {c_i^{F}  K }\ . \equn
$$

\section{Box Coefficients of The Six Point $N=1$ Amplitudes}

We can organise the six point amplitudes according to the number of
negative helicities; amplitudes with zero, one, five or
six vanish in any supersymmetric theory.
The amplitudes with two negative helicities are the MHV amplitudes, which
 were computed previously~\cite{BDDKb}, while those
with four are 
the ``googly'' MHV amplitudes which are obtained by conjugation of the MHV amplitudes.    
Here we present the remaining box coefficients and examine the twistor structure of
all the six point amplitudes.

The two independent types of six point amplitude have rather different box
structures.  The MHV amplitudes contain ``two-mass easy'' and single
mass boxes, whereas the amplitudes with three negative helicities contain 
``two-mass hard'' and single mass boxes. This feature does not extend to higher point functions.

\noindent
{\bf MHV Amplitudes }

\nopagebreak
There are three independent MHV amplitudes. In terms of the $D=6$ boxes the box parts of 
these amplitudes are , 
$$
\eqalign{
A(1^-,2^-,3^+,4^+,5^+,6^+)|_{{\rm box}} =& \; 0
\cr
A(1^-,2^+,3^-,4^+,5^+,6^+)|_{{\rm box}} =&
\; b^{D=6}_1 \Ieasy{3}+b^{D=6}_2\Ione{5}+b^{D=6}_3\Ione{3}
\cr
A(1^-,2^+,3^+,4^-,5^+,6^+)|_{{\rm box}} =&
\; c^{D=6}_1 \Ieasy{1}+c^{D=6}_2 \Ieasy{3}+ c^{D=6}_3\Ione{6}+c^{D=6}_4\Ione{3}
\cr}
\equn
$$

where,
$$
\eqalign{
b_1^{D=6}
= \Atreemhv_{13}{ \tr_+(1325)\tr_+(1352) \over s_{13}^2s_{25} }
\;\;\;& \;\;\;
b_2^{D=6} 
= \Atreemhv_{13}  { \tr_+(1324)\tr_+(1342) \over s_{13}^2s_{24} }
\cr
b_3^{D=6} 
=  \Atreemhv_{13} & {   \tr_+(1326)\tr_+(1362) \over s_{13}^2s_{26} }
\cr} \equn
$$

$$
\eqalign{
c_1^{D=6}= \Atreemhv_{14}  { \tr_+(1436)\tr_+(1463) \over s_{14}^2s_{36} }
\;\; & \;\;
c_2^{D=6}= \Atreemhv_{14}  { \tr_+(1425)\tr_+(1452) \over s_{14}^2s_{25} }
\cr
c_3^{D=6}= \Atreemhv_{14}  { \tr_+(1435)\tr_+(1453) \over s_{14}^2s_{35} }
\;\; & \;\;
c_4^{D=6}= \Atreemhv_{14}  { \tr_+(1426)\tr_+(1462) \over s_{14}^2s_{26} }
\cr} 
\equn
$$
where $\tr_+(\rm abcd)=\spb{a}.b\spa{b}.c\spb{c}.d\spa{d}.a$.
If we examine the coefficients of the $F$-functions we have, for example,
$$
b_1^{F}
 = \Atreemhv_{13} \times { \tr_+(1325)\tr_+(1352) \over s_{13}^2s_{25}^2 }
=\Atreemhv_{13}\times {
\spa3.2 \spa1.5 \spa3.5\spa2.1 \over
\spa1.3^2 \spa2.5^2 },
\equn
$$
which is a holomorphic function ({\it i.e}. a function of $\lambda$ alone).
The amplitude has an overall factor in dimensional regularisation of 
$ r_\Gamma$, where
$$
r_\Gamma\ ={(\mu^2)^{\eps} \over (4 \pi)^{2-\e}}
{\Gamma(1+\e)\Gamma^2(1-\e)\over\Gamma(1-2\e)} \,,
\equn
\label{cGamma}
$$
which we will not write explicitly here or in following cases.

\vskip 3truemm
\noindent
{\bf Amplitudes with three minus helicities}

There are also three independent amplitudes with three minus helicities: 
$A(1^-,2^-,3^-,4^+,5^+,6^+)$, $A(1^-,2^-,3^+,4^-,5^+,6^+)$
and $A(1^-,2^+,3^-,4^+,5^-,6^+)$. Of these,  the first consists only
of triangle and bubble integrals~\cite{BBDD} so we have a trivial
box structure,
$$
\eqalign{
A(1^-,2^-,3^-,4^+,5^+,6^+)|_{{\rm box}} =& \; 0.
\cr}\equn
$$
The next  amplitude, $A(1^-,2^-,3^+,4^-,5^+,6^+)$, 
does have a non-trivial box structure, which we express in terms of $D=6$ boxes as,
$$
\eqalign{
A(1^-,2^-,3^+,4^-,5^+,6^+)|_{{\rm box}} =&
\;
c_1^{D=6} \Ihard{2;6}+c_2^{D=6}\Ihard{2;2}+c_3^{D=6}\Ihard{2;4}
+c_4^{D=6}\Ione{5}+c_5^{D=6}\Ione{6},
\cr}\equn
$$
where the integral boxes are,

\vskip 0.5 truecm

\begin{picture}(90,55)(0,0)

\Text(5,25)[c]{$c_1^{D=6}$}
\Line(20,10)(60,10)
\Line(20,40)(60,40)
\Line(20,10)(20,40)
\Line(60,40)(60,10)

\Line(20,10)(20,0)
\Line(60,10)(60,0)

\Line(20,40)(15,50)
\Line(20,40)(25,50)

\Line(60,40)(55,50)
\Line(60,40)(65,50)

\Text(8,0)[l]{\small $5$}
\Text(3,50)[l]{\small $6$}
\Text(28,50)[l]{\small $1$}

\Text(48,50)[l]{\small $2$}
\Text(68,50)[l]{\small $3$}

\Text(63,0)[l]{\small $4$}

\end{picture}
\begin{picture}(90,55)(0,0)
\Text(0,25)[c]{$+c_2^{D=6}$}
\Line(20,10)(60,10)
\Line(20,40)(60,40)
\Line(20,10)(20,40)
\Line(60,40)(60,10)

\Line(20,10)(20,0)
\Line(60,10)(60,0)

\Line(20,40)(15,50)
\Line(20,40)(25,50)

\Line(60,40)(55,50)
\Line(60,40)(65,50)

\Text(8,0)[l]{\small $1$}
\Text(3,50)[l]{\small $2$}
\Text(28,50)[l]{\small $3$}

\Text(48,50)[l]{\small $4$}
\Text(68,50)[l]{\small $5$}

\Text(63,0)[l]{\small $6$}

\end{picture}
\begin{picture}(90,55)(0,0)
\Text(0,25)[c]{$+c_3^{D=6}$}
\Line(20,10)(60,10)
\Line(20,40)(60,40)
\Line(20,10)(20,40)
\Line(60,40)(60,10)

\Line(20,10)(20,0)
\Line(60,10)(60,0)

\Line(20,40)(15,50)
\Line(20,40)(25,50)

\Line(60,40)(55,50)
\Line(60,40)(65,50)

\Text(8,0)[l]{\small $3$}
\Text(3,50)[l]{\small $4$}
\Text(28,50)[l]{\small $5$}

\Text(48,50)[l]{\small $6$}
\Text(68,50)[l]{\small $1$}

\Text(63,0)[l]{\small $2$}

\end{picture}
\begin{picture}(90,55)(0,0)
\Text(0,25)[c]{$+c_4^{D=6}$}
\Line(20,10)(60,10)
\Line(20,40)(60,40)
\Line(20,10)(20,40)
\Line(60,40)(60,10)

\Line(20,10)(20,0)
\Line(60,10)(60,0)

\Line(20,40)(20,50)

\Line(60,40)(55,50)
\Line(60,40)(65,50)

\Text(8,0)[l]{\small $3$}
\Text(8,50)[l]{\small $4$}

\Text(48,50)[l]{\small $5$}
\Text(57,50)[l]{\small $6$}
\Text(68,50)[l]{\small $1$}

\Text(63,0)[l]{\small $2$}

\end{picture}
\begin{picture}(90,55)(0,0)
\Text(0,25)[c]{$+c_5^{D=6}$}
\Line(20,10)(60,10)
\Line(20,40)(60,40)
\Line(20,10)(20,40)
\Line(60,40)(60,10)

\Line(20,10)(20,0)
\Line(60,10)(60,0)

\Line(20,40)(20,50)

\Line(60,40)(55,50)
\Line(60,40)(65,50)

\Text(8,0)[l]{\small $4$}
\Text(8,50)[l]{\small $5$}

\Text(48,50)[l]{\small $6$}
\Text(57,50)[l]{\small $1$}
\Text(68,50)[l]{\small $2$}

\Text(62,0)[l]{\small $3$}

\end{picture}

and the coefficients have been computed to be,

$$
\begin{array}{llll}
c_1^{D=6}
&=&\displaystyle i { ({ \langle 3 | \Slash{P} | 1 \rangle })^2
{ \langle 5 | \Slash{P} | 4  \rangle }
{ \langle 3 | \Slash{P} | 5 \rangle }
\over
{ \langle 4  | \Slash{P} | 5 \rangle }
{ \langle 2  | \Slash{P} | 5 \rangle }   }
{ \spa5.1
\over \spb2.3\spa5.6 \spa6.1 P^2 },
&\hskip 1.0cm P=P_{234},
\cr
c_2^{D=6}
&=& \displaystyle i
 { ({ \langle 3 | \Slash{P} | 4  \rangle })^2
{ \langle 6 | \Slash{P} | 1  \rangle }
\over { \langle 1  | \Slash{P} | 6 \rangle }   }
{ \spb3.1 \spa6.4
\over \spb1.2\spb2.3\spa4.5\spa5.6 P^2  },
&\hskip 1.0cm P=P_{123},
\cr
c_3^{D=6}
&=& \displaystyle i { ({ \langle 6 | \Slash{P} | 4 \rangle })^2
{ \langle 2 | \Slash{P} | 4  \rangle }
{ \langle 3 | \Slash{P} | 2 \rangle }
\over
{ \langle 2  | \Slash{P} | 3 \rangle }
{ \langle 2  | \Slash{P} | 5 \rangle }   }
{ \spb6.2
\over \spa4.5 \spb6.1 \spb1.2 P^2 },
&\hskip 1.0cm P=P_{345},
\cr
c_4^{D=6}
&=&\displaystyle i
{ ( { \langle 3 | \Slash{P} |  1 \rangle } )^2
{ \langle 2 | \Slash{P} | 1 \rangle }
\over
{ \langle 2  |\Slash{P} | 5 \rangle }
}
{ \spa2.4
\over \spa5.6 \spa6.1  P^2
\spb2.4  },
&\hskip 1.0cm P=P_{234},
\cr
c_5^{D=6}
&=&\displaystyle i
{ ( { \BRR64 } )^2
{ \langle 6 | \Slash{P} | 5 \rangle }
\over
{ \langle 2  |\Slash{P} | 5 \rangle }
}
{ \spb3.5
\over \spb6.1 \spb1.2  P^2
\spa3.5  },
&\hskip 1.0cm P=P_{345},
%\cr}
\end{array}
\equn
$$
where $\la a | \Slash{K} | c \ra\equiv \la a^+ | \Slash{K} | c^+ \ra$. 

The remaining amplitude, $A(1^-,2^+,3^-,4^+,5^-,6^+)$, contains all
six one-mass and all six ``two-mass-hard'' boxes,
$$
\eqalign{
A(1^-,2^+,  3^-,4^+, & 5^-, 6^+)_{\rm box}= \cr
&
a_{1}^{D=6}\Ione{4}
+a_{2}^{D=6}\Ione{5}
+a_{3}^{D=6}\Ione{6}
+a_{4}^{D=6}\Ione{1}
+a_{5}^{D=6}\Ione{2}
+a_{6}^{D=6}\Ione{3}
\cr
+&
b_1^{D=6} \Ihard{2;5}
+b_2^{D=6} \Ihard{2;6}
+b_3^{D=6} \Ihard{2;1}
+b_4^{D=6} \Ihard{2;2}
+b_5^{D=6} \Ihard{2;3}
+b_6^{D=6} \Ihard{2;4}.
\cr} \equn
$$
Fortunately these are not all independent and symmetry demands relationships
amongst the $a^{D=6}_i$'s,
$$
\eqalign{
a^{D=6}_3(123456)=a^{D=6}_1(345612), \;\;\;
a^{D=6}_5(123456)=a^{D=6}_1(561234), \;\;\;
\cr
a^{D=6}_4(123456)=a^{D=6}_2(345612), \;\;\;
a^{D=6}_6(123456)=a^{D=6}_2(561234), \;\;\;
\cr
a^{D=6}_2(123456) = \bar{a}^{D=6}_1(234561), \;\;\;
a^{D=6}_1(123456)=a^{D=6}_1(321654),\; \;\;\;
\cr}\equn
$$
where $\bar a^{D=6}_1$ denotes $a^{D=6}_1$ with $\spa{i}.j \leftrightarrow \spb{i}.j$.
Thus there is a single independent $a^{D=6}_i$. Similarly we can use symmetry to
generate all the $b^{D=6}_i$'s from $b^{D=6}_1$.  The
expressions for $a^{D=6}_1$ and $b^{D=6}_1$ are, 
$$
\begin{array}{llll}
a^{D=6}_1 & = &
\displaystyle  i{ \BRR{2}{5}^2 \BRR15 \BRR35 \over \BRR36 \BRR14 P^2  }
{ \spa3.1 \over \spb1.3 \spa4.5 \spa5.6 },
&\hskip 0.5 truecm
P=P_{123},
\cr
b^{D=6}_1 &= &
\displaystyle i{ \BRR{2}{5}^2  \BRR35 \BRR24  \BRR43 \over \BRR36 \BRR14 \BRR34  P^2  }
{ 1 \over \spb1.2 \spa5.6 },
&\hskip 0.5 truecm
P=P_{123}.
%\cr}
\end{array}
\equn
$$

\noindent
{\bf Googly MHV Amplitudes}

The googly MHV amplitudes can be obtained from the MHV amplitudes by conjugation. 
These amplitudes are useful for testing hypotheses regarding
amplitudes containing four minus helicities. For example we have, 
$$
A(1^+,2^-,3^+,4^-,5^-,6^-)|_{{\rm box}} =
b^{D=6}_1{}' \Ieasy{2;3}+b^{D=6}_2{}'\Ione{5}+b^{D=6}_3{}'\Ione{3},
\equn
$$
with $b^{D=6}_i{}' =\overline{b}^{D=6}_i$. The coefficients of the $F$-functions are 
anti-holomorphic functions, {\it e.g.}
$$
b_1^F{}'
=  \bar{A}^{\rm tree}_{24} \times {
\spb3.2 \spb1.5 \spb3.5\spb2.1 \over
\spb1.3^2 \spb2.5^2 }.
\equn
$$

\section{Cut Constructibility }

In this section, we review and discuss the status of the box
coefficients calculated by evaluating the cuts in one-loop amplitudes. 
In ref~\cite{BDDKa}
the concept that an amplitude was ``cut constructible'' was
introduced. At first sight the meaning of this term appears
obvious: that one may calculate an amplitude from a knowledge of
its cuts,  
$$
C_{i\ldots j}  \equiv  
{ i \over 2} \int \dlips\biggl[ A^{\rm tree}(\ell_1,i,i+1,\ldots,
j,\ell_2) \times A^{\rm tree}(-\ell_2,j+1,j+2,\ldots,i-1,-\ell_1)
\biggr] 
\equn
$$
in all channels $i\ldots j$.

Any amplitude involving massless
particles can be reconstructed from a {\it full} knowledge of its
cuts (see ref.~\cite{Bern:2004cz} for a modern review).  
This means that if we calculate
the cuts precisely and regularize them in the same fashion as
the amplitude, then  we can determine any amplitude. Specifically, if we
regularize the amplitude by dimensional regularization then, for consistency,
in the
cut $C_{i\ldots j}$ we should use tree amplitudes with
external momenta in four dimensions, while the  momenta
crossing the cut should reside in $4-2\eps$ dimensions. 
These are not the normal tree amplitudes. 
In ref~\cite{DimShift} this was explicitly realized and used to
determine a specific non-supersymmetric amplitude. This method may also
be used for amplitudes beyond one-loop~\cite{BRY,BDDPR}.

Fortunately, for $N=4$ and $N=1$ supersymmetric gauge theory
amplitudes it is not necessary to evaluate the cuts in this
precise manner, instead  one may calculate the cut using amplitudes
where the cut legs lie in four dimensions. This means that the cut can be
evaluated using the conventional four dimensional tree amplitudes.  In principle this 
introduces errors in the trees at $O(\eps)$.  It is
non-trivial that these errors do not produce finite terms within
the possibly divergent integrals.  The proof of this 
lies in a detailed study of the possible integral functions which
may occur within a one-loop calculation. For the restricted case
of supersymmetric theories the cuts contain enough information to
determine the coefficients of these functions unambiguously.  This
is a more precise definition of `cut-constructibility'.  A small
number of supergravity amplitudes are also
cut-constructible~\cite{DunNor}. For $N=4$
amplitudes the integral functions are precisely scalar boxes. For
$N=1$ amplitudes we have scalar boxes plus scalar triangles and
bubbles.  As presented in refs~\cite{BDDKa,BDDKb} the uniqueness
of the coefficients hinges on the uniqueness of the classes of
logarithms appearing in the cuts.  Generically, the boxes are in a
different class of functions from bubbles and triangles since the
latter do not contain terms like
$$
\ln (P^2_{i\ldots j}) \ln(P^2_{i'\ldots j'}), \equn
$$ 
which reside in boxes.  By considering such terms, or
specifically the coefficients of $\ln (P^2_{i'\ldots j'})$ in the
$P_{i\ldots j}$-channel cut, together with the limited number of boxes
which may contain such a term in the $P_{i\ldots j}$-channel, one can
show that the coefficients are uniquely defined.  Generically this
makes it unambiguous to extract the coefficients of boxes from a
single cut.  In performing a cut in, {\it e.g.,} the $P_{i\ldots
j}$-channel we can determine boxes with terms like $\ln
(P_{i\ldots j}^2) \ln(P_{i'\ldots j'}^2)$ by determining the
coefficients of $\ln(P_{i'\ldots j'}^2)$ in this cut.  
In fact we tend not to evaluate the cut directly, but
rather manipulate the cut into a form where it can be recognized as
the cut of specific scalar boxes with coefficients. 

We have carried out such a process in evaluating the
coefficients of the box functions in the six point amplitudes. We shall
illustrate this explicitly in the following section where we evaluate the
coefficients of certain boxes in higher point functions.

\section{Higher Point Box Coefficients}

In this section we evaluate some sample box coefficients for certain
$n$-point amplitudes. This will enable us to examine whether the 
twistor
space structure of the six-point amplitudes extends to higher
point amplitudes.

For higher point amplitudes the number of helicity configurations grows
quite rapidly with  
increasing numbers
of legs.  As our first example we will consider the specific amplitude,
$$
A^{N=1,\rm chiral}(1^- 2^- \cdots  j^+ (j+1)^- 5^+ \cdots n^+)\ . \equn
$$ We calculate the $123\cdots j$-cut of this amplitude, {\it i.e.},
$$
C_{123\cdots j} = {i \over 2} \int \dlips \sum_{h\in \{-1/2,0,1/2\}}
\Atree(\ell_1^h,1^-,2^-,\cdots,j^+,\ell_2^{-h})
\Atree((-\ell_2)^h, (j+1)^-,\cdots, n^+,(-\ell_1)^{-h}) \,,\equn
$$
The sum is over the particles in the $N=1$ chiral multiplet.
The two tree amplitudes are a MHV
amplitude and a MHV-googly amplitude.  For MHV amplitudes the
different tree amplitudes for different particle types
are related by simple cofactors
determined by solving the supersymmetric Ward
identities~\cite{SWI,ManganoReview}. Using these to replace the
tree amplitudes by the amplitudes for scalars with cofactors and
summing the cofactors we obtain, 
$$  
\eqalign{
C_{123\cdots j} = {i \over 2} \int \dlips  
&
\Atreemhv
(\ell_1^s,1^-,2^-,\cdots,j^+,\ell_2^s)
\cr
&
\times
{A}^{\rm tree\ MHV \ googly}((-\ell_2)^s,
(j+1)^-,\cdots, n^+,(-\ell_1)^{s}) \times \rho^{N=1} } \  \equn
$$
where, 
$$
\rho^{N=1}= -x +2 -{1\over x} = -{  (x-1)^2 \over x } \ , 
\;\;\;\;
{\rm with}
\;\;\;
x= { \spb j.{\ell_2} \spa{j+1}.{\ell_2} \over  \spb j.{\ell_1} \spa {j+1}.{\ell_1} }
$$
so that, 
$$
\eqalign{
\rho^{N=1} &= -{ \spb j.{\ell_1} \spa {j+1}.{\ell_1}\over\spb j.{\ell_2} \spa{j+1}.{\ell_2} }
\left(
{ \spb j.{\ell_2} \spa{j+1}.{\ell_2} \over  \spb j.{\ell_1} \spa{j+1}.{\ell_1} } -1 \right)^2 
=-{  \la  j | P_{123\cdot j} | {j+1} \ra^2
\over  \spb j.{\ell_1} \spa{j+1}.{\ell_1}\spb j.{\ell_2} \spa{j+1}.{\ell_2}  }
\cr}\ . \equn
$$
This gives the integrand above as,
$$
\eqalign{&
{ \spb j .{\ell_1}^2\spb j.{\ell_2}^2
\over \spb1.2\spb2.3\cdots\spb {j-1}. j \spb j .{\ell_2}\spb{\ell_2}.{\ell_1}\spb{\ell_1}.1 }
\times
{ \spa{j+1}.{\ell_1}^2\spa {j+1}.{\ell_2}^2
\over \spa{j+1}.{j+2}\spa{j+2}.{j+3}\cdots \spa{n-1}.n \spa{n}.{\ell_1}\spa{\ell_1}.{\ell_2}\spa{\ell_2}.{j+1} }\cr
&\times
{  \la j | P_{123\cdots j} | j+1 \ra^2
\over  \spb j.{\ell_1} \spa{j+1}.{\ell_1}\spb j .{\ell_2} \spa{j+1}.{\ell_2}  } \cr&
={  \la j | P_{123\cdots j} | j+1 \ra^2  \over \spb1.2\spb2.3\cdots\spb {j-1}.j \spa{j+1}.{j+2}\spa{j+2}.{j+3}\cdots \spa{n-1}.n
P^2_{123\cdots j}}
\times
{ \spb j.{\ell_1}
\over \spb{\ell_1}.1 }
\times
{ \spa j+1.{\ell_1}
\over \spa{n}.{\ell_1} }
\cr
& ={  \la j | P_{123\cdots j} | j+1 \ra^2  \over \spb1.2\spb2.3\cdots\spb {j-1}.j \spa{j+1}.{j+2}\spa{j+2}.{j+3}\cdots \spa{n-1}.n
P^2_{123}}
\times
{ \spb j .{\ell_1}\spa{\ell_1}.1 \spa {j+1}.{\ell_1}\spb{\ell_1}.n
\over (\ell_1-k_1)^2 (\ell_1+k_n)^2  }
\ . \cr} \equn
 $$
This corresponds to the cut of a box integral with 
integrand quadratic in the loop momentum,
{\it i.e.,}
$$
\eqalign{
C_{123\cdots j} &=  {  \la j | P_{123\cdots j} | j+1 \ra^2  \over
\spb1.2\cdots\spb {j-1}.j \spa{j+1}.{j+2}\spa{j+2}.{j+3}\cdots
\spa{n-1}.n P^2_{123\cdots j}} \cr & \null\hskip 1.0 truecm \times \left( I_2^{2m h} [
\spb j.{\ell_1}\spa{\ell_1}.1 \spa j+1.{\ell_1}\spb{\ell_1}.n ] \right)_{\rm cut} \; . 
\cr}\equn
$$
The specific box integral is the ``two mass hard'' depicted below,
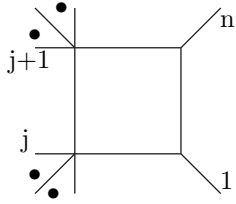
\begin{center}
\begin{picture}(120,95)(0,0)
% Picture 3 2 mass hard
% Lines of Box
\Line(30,20)(70,20)
\Line(30,60)(70,60)
\Line(30,20)(30,60)
\Line(70,60)(70,20)

%Lines of corner lines
\Line(30,20)(15,5)
\Line(30,60)(15,75)
\Line(70,20)(85,5)
\Line(70,60)(85,75)

%Massive Lines
\Line(30,20)(30,5)
\Line(30,20)(15,20)
\Text(22,5)[c]{\small$\bullet$}
\Text(15,12)[c]{\small $\bullet$}

%Massive Lines
\Line(30,60)(30,75)
\Line(30,60)(15,60)
\Text(15,65)[c]{\small$\bullet$}
\Text(25,75)[c]{\small $\bullet$}

\Text(10,25)[l]{\small $\hbox{\rm j}$}
\Text(85,10)[l]{\small $\hbox{\rm 1}$}
\Text(85,70)[l]{\small $\hbox{\rm n}$}
\Text(5,55)[l]{\small $\hbox{\rm j+1}$}

%\Text(40,40)[l]{$I^{2mh}_{4:r;i}$}
\end{picture}
\end{center}
with a non-trivial (quadratic in loop momenta) numerator.

Rewriting
the numerator,
$$
\la j | \Slash{\ell_1} | 1 \ra \la n | \Slash{\ell_1} | j+1\ra
= { \la j^+ | \Slash{\ell_1} | 1^+ \ra
\la 1^+ | P | n^+ \ra \la n | \Slash{\ell_1} |j+1\ra
\over \la 1^+ | P | n^+ \ra}
={ \la j | \Slash{\ell_1} \Slash{ 1}  \Slash{P}
\Slash{k_n}  \Slash{\ell_1} |j+1\ra
\over \la 1 |\Slash{P} | n \ra}\, ,  \equn
$$
and commuting the cut momenta toward $\Slash P = \Slash \ell_1-\Slash\ell_2$,
$$
\eqalign{
\Slash{\ell_1} \Slash{ 1}  \Slash{P}
\Slash{k_n}  \Slash{\ell_1}
&= (2\ell_1\cdot k_1) \Slash{P}   \Slash{k_n}  \Slash{\ell_1}
- \Slash{ 1}\Slash{\ell_1} \Slash{P}
\Slash{k_n}  \Slash{\ell_1}
\cr
& =(2\ell_1\cdot 1) \Slash{P}   \Slash{k_n}  \Slash{\ell_1}
-(2\ell_1\cdot k_n)\Slash{ 1}\Slash{\ell_1} \Slash{P}
+\Slash{ 1}\Slash{\ell_1} \Slash{P}
  \Slash{\ell_1}\Slash{k_n}
\cr
&=-(\ell_1-k_1)^2 \Slash{P}   \Slash{k_n}  \Slash{\ell_1}
-(\ell_1+k_n)^2 \Slash{ k_1}\Slash{\ell_1} \Slash{P}
+(2\ell_1\cdot P) \Slash{ k_1}
  \Slash{\ell_1}\Slash{k_n}
 \ . \cr}\,\equn
$$ 
In this expression the first two terms cancel a propagator 
yielding triangle integrals - which we discard for the present
purposes - and the third term can be rearranged as $(2\ell_1\cdot P)
=-(\ell_1-P)^2+\ell_1^2+P^2=-\ell_2^2+\ell_1^2+P^2 \equiv P^2$ discarding momenta
null on the cut. 
The remaining expression is a box with linear integrand which can be
evaluated and the result expressed as a $D=6$ scalar box function,
$$
C_{123\cdots j} = 
{  \la j  | P_{123\cdots j} | (j+1) \ra^2 \la n |\Slash{P} | 1 \ra  \spb1.j \spa{j+1}.n
\over
\la 1 |\Slash{P} | n \ra \spb1.2\spb2.3 \cdots\spb {j-1}.j 
\spa{j+1}.{j+2}\spa{j+2}.{j+3}\cdots \spa{n-1}.n
P^2_{123\cdots j}}  \left(  I_4^{2m h , D=6}  \right)_{\rm cut} \equn
$$
so we deduce, using the arguments of the previous section,  that the coefficient of the box is
$$
f_1^{D=6}=
i {  \la j  | \Slash{P} | (j+1) \ra^2 \la n |\Slash{P} | 1 \ra  \spb1.j \spa{j+1}.n
\over
\la 1 |\Slash{P} | n \ra \spb1.2\spb2.3 \cdots\spb {j-1}.j 
\spa{j+1}.{j+2}\spa{j+2}.{j+3}\cdots \spa{n-1}.n
P^2}
\; ,
\;\;\; P=P_{123\cdots j} \; ,
\equn 
$$
which is a generalisation of the coefficient $c_2$ within the six point amplitude
$A(1^-,2^-,3^+,4^-,5^+,6^+)$. 

As a further example, by looking at the $C_{n\cdots j-1}$ cut we can deduce that the amplitude,
$$
A^{N=1, \rm chiral}(1^- 2^- \cdots  (j-1)^- j^+ (j+1)^+ \cdots k^- \cdots (n-1)^+ n^+)\ , \equn
$$
(where legs $1$ to $j-1$ and leg $k$ have negative helicity and the remainder have positive helicity)
contains  boxes, 
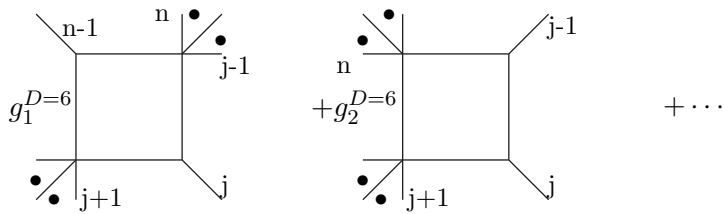
\begin{center}
%\begin{picture}(120,95)(0,0)
%\Text(5,45)[l]{
%$A(1^- 2^-\cdots  n^+)=$}
%\end{picture}
\begin{picture}(120,95)(0,0)
%Picture 2: 2 mass easy

\Text(15,40)[c]{ $g_1^{D=6}$}

% Lines of Box
\Line(30,20)(70,20)
\Line(30,60)(70,60)
\Line(30,20)(30,60)
\Line(70,60)(70,20)

%Lines of corner lines
\Line(30,20)(15,5)
\Line(30,60)(15,75)
\Line(70,20)(85,5)
\Line(70,60)(85,75)

%Massive Lines
\Line(30,20)(30,5)
\Line(30,20)(15,20)
\Text(22,5)[c]{\small$\bullet$}
\Text(15,12)[c]{\small $\bullet$}

%Massive Lines
\Line(70,60)(70,75)
\Line(70,60)(85,60)
\Text(75,75)[c]{\small$\bullet$}
\Text(85,65)[c]{\small $\bullet$}

\Text(85,55)[l]{\small $\hbox{\rm j-1}$}
\Text(60,75)[l]{\small $\hbox{\rm n}$}

\Text(32,5)[l]{\small $\hbox{\rm j+1}$}
\Text(85,10)[l]{\small $\hbox{\rm j}$}
\Text(25,70)[l]{\small $\hbox{\rm n-1}$}

% \Text(40,40)[l]{$I^{2me}_4$}
\end{picture}
\begin{picture}(120,95)(0,0)
% Picture 3 2 mass hard
\Text(10,40)[c]{ $+g_2^{D=6}$}
% Lines of Box
\Line(30,20)(70,20)
\Line(30,60)(70,60)
\Line(30,20)(30,60)
\Line(70,60)(70,20)

%Lines of corner lines
\Line(30,20)(15,5)
\Line(30,60)(15,75)
\Line(70,20)(85,5)
\Line(70,60)(85,75)

%Massive Lines
\Line(30,20)(30,5)
\Line(30,20)(15,20)
\Text(22,5)[c]{\small$\bullet$}
\Text(15,12)[c]{\small $\bullet$}

%Massive Lines
\Line(30,60)(30,75)
\Line(30,60)(15,60)
\Text(15,65)[c]{\small$\bullet$}
\Text(25,75)[c]{\small $\bullet$}

\Text(32,5)[l]{\small $\hbox{\rm j+1}$}
\Text(85,10)[l]{\small $\hbox{\rm j}$}
\Text(85,70)[l]{\small $\hbox{\rm j-1}$}
\Text(5,55)[l]{\small $\hbox{\rm n}$}

% \Text(40,40)[l]{$I^{2mh}_4$}
\end{picture}
\begin{picture}(80,95)(0,0)
\Text(5,40)[l]{$+\cdots $}
\end{picture}

\end{center}

The first appearance of the two-mass easy box in non-MHV amplitudes occurs at seven point amplitudes. 
The coefficients are
$$
\eqalign{
g_1^{D=6} =& 
-i{ \BRR{n}{k}^2 \BRR{n}{n-1} 
 \spa{k}.{n-1} \spb {n-1}.j \spa{j}.k 
\over
 \spb{n}.1\spb1.2 \cdots \spb{j-2}.{j-1}
\spa{j}.{j+1} \spa{j+1}.{j+2} \cdots \spa{n-2}.{n-1}
\BRR{j-1}{n-1}
\spa{n-1}.j P^2 
}
\cr
g_2^{D=6} =& 
i{\BRR{n}{k}^2  \BRR{j-1}{k}  \BRR{j}{j-1} \spb{n}.{j-1} \spa{j}.{k} 
\over \spb{n}.1\spb1.2 \cdots \spb{j-2}.{j-1}
\spa{j}.{j+1} \spa{j+1}.{j+2} \cdots \spa{n-2}.{n-1}
\BRR{j-1}{j} \BRR{j-1}{n-1} P^2   
}
\cr}
\equn
$$
Using symmetry arguments various
other box coefficients can be obtained from these expressions by relabeling.

\section{Twistor Structure}

It was observed by Witten~\cite{Witten:2003nn} that 
the twistor space properties of amplitudes expressed
in terms of the helicity states ($\lambda_i, \tilde\lambda_i$)
can be investigated using particular differential operators. 
Specifically, if a
function has non-zero support when points $i$, $j$ and $k$ are
{\it collinear} in twistor space, then it is
annihilated by the operator
$$
[F_{ijk} , \eta ] =
 \spa{i}.j \left[{ \partial\over \partial\tilde\lambda_k},\eta\right]
+\spa{j}.k \left[{ \partial\over \partial\tilde\lambda_i},\eta\right]
+\spa{k}.i \left[{ \partial\over \partial\tilde\lambda_j},\eta\right]
\ , 
\equn
$$
where the square brackets indicate spinor products rather than commutators. 
Similarly, annihilation by the operator
$$
\eqalign{K_{ijkl}=\frac14\Big[ \la {i j} \ra \epsilon^{ {\dot a
\dot b}} \frac{\partial}{\partial{\tilde \lambda}^{{\dot
a}}_{{k}}} \frac{\partial}{\partial {\tilde \lambda}^{{\dot
b}}_{{l}}} &-\la {i k} \ra {\epsilon}^{{\dot a \dot b}}
\frac{\partial}{\partial {\tilde \lambda}^{{\dot a}}_{{j}}}
\frac{\partial}{\partial {\tilde \lambda}^{{\dot b}}_{{l}}} +\la
{i l} \ra {\epsilon}^{{\dot a \dot b}} \frac{\partial}{\partial
{\tilde \lambda}^{{\dot a}}_{{j}}} \frac{\partial}{\partial
{\tilde \lambda}^{{\dot b}}_{{k}}} \cr & +\la {j k} \ra
{\epsilon}^{{\dot a \dot b}} \frac{\partial}{\partial {\tilde
\lambda}^{{\dot a}}_{{i}}} \frac{\partial}{\partial {\tilde
\lambda}^{{\dot b}}_{{l}}} +\la {j l} \ra {\epsilon}^{{\dot a \dot
b}} \frac{\partial}{\partial{\tilde \lambda}^{{\dot a}}_{{k}}}
\frac{\partial}{\partial {\tilde \lambda}^{{\dot b}}_{{i}}} -\la
{k l} \ra {\epsilon}^{{\dot a \dot b}} \frac{\partial}{\partial
{\tilde \lambda}^{{\dot a}}_{{j}}} \frac{\partial}{\partial
{\tilde \lambda}^{{\dot b}}_{{i}}} \Big]} \ , \equn
$$ 
indicates {\it co-planarity} of points $i,j,k$ and $l$ in twistor
space.

Here we will explore the twistor space structure of the box
coefficients of the $N=1$ amplitudes.  At tree-level an important
implication of the CSW-formalism is that the twistor space properties
of amplitudes are completely determined by the number of minus legs.
For this reason we organise the one-loop amplitudes according to the
number of negative helicities.
We have investigated the twistor space properties for all the possible
5-point box coefficients and all the 6-point box coefficients
together with the $n$-point coefficients of the previous section. This
was carried out by generating sets of on-shell kinematic points
consisting of specific values of $\lambda_i$ and $\tilde{\lambda}_i$
and determining the action of the operators at these points.

For the six point amplitudes there are three
different classes of amplitudes organised by the number
of negative helicities: MHV-amplitudes, next-to-MHV amplitudes and googly MHV-amplitudes.
For the $n$-point  amplitudes we have extended certain six point amplitudes 
by adding extra plus legs to the MHV side of the cut and extra minus legs to
the googly side. This produces the following classes of $n$-point
configurations: $(-\cdots-+\cdots+-+\cdots+)$ and $(-\cdots - +
- + \cdots +)$.

For the MHV-amplitudes 
all helicity configurations for the box coefficients are holomorphic and
are thus  annihilated by any
$F_{ijk}$ and $K_{ijkl}$ operator, as noted in~\cite{Cachazo:2004zb}. 
The geometric picture of these
configurations is simply a line in twistor space.

Now we consider next-to-MHV amplitudes with three minus helicities.
By acting with the $K_{ijkl}$ operators we find that the
box coefficients are annihilated for any four points,
$$
K_{ijkl}\biggl[ c^{F}_ {\rm next\ to \ MHV}\biggr] =0,
\equn
$$
indicating a geometric picture where all points lie in a plane in twistor space.

The line structure of the box coefficients can be deduced by acting
with the $F_{ijk}$ operators. In the cuts we have used to determine
these coefficients, there is a MHV tree amplitude on one side of the
cut (the ``mostly plus side'') and a googly MHV tree amplitude on the
other (the ``mostly minus side'').  The box coefficients calculated
from each cut will be annihilated by $F_{ijk}$ when $i$, $j$ and $k$
are any legs lying on the mostly plus side of that cut, indicating
that these legs define points in twistor space that lie on a line.
Similar behaviour was found for the box coefficients in $N=4$
amplitudes~\cite{BeDeDiKo,{Britto:2004tx}}.

For the $q(>3)$ minus configurations, the box coefficients are only
annihilated by $F_{ijk}$ operators where all three of the points lie
on the MHV, mostly plus, side of the cut used to calculate them.
These points will lie on a line in twistor space.  Hence the box
coefficients are annihilated by any $K_{ijkl}$ operator where three or
more of these points lie on the line.  For generic points in twistor
space, we have confirmed explicitly that only these $K_{ijkl}$
operators annihilate the box coefficients.  The geometric
interpretation is thus of $n-q$ points lying on a line with no
restriction on the positions of the remaining $q$ points.
In general, if a box has a cut in the channel $C_{i\ldots j}$ and
$\Atree({i\ldots j})$ is a MHV tree amplitude, then the
box coefficient is supported on configurations in twistor space where
points ${i\ldots j}$ are collinear.  If there are two or more such
cuts, this would imply a support of two or more lines with the
remaining points unrestricted.  When any pair of these cuts have a
common leg, the corresponding lines intersect at the common point.

We have presented explicitly the results for the $N=1$ chiral multiplet. Since the 
$N=1$ vector multiplet is a linear combination of this and the $N=4$ multiplet,
the box coefficients of the $N=1$ vector multiplet will also have planar support
for next to MHV amplitudes.

\section{Conclusions}

In the twistor space realisation of gauge theory amplitudes many
fascinating geometric features appear. These are of interest both
formally and, possibly, practically in the determination of scattering
amplitudes.  One-loop amplitudes can be expressed as a sum of integral
functions whose coefficients, in
particular the coefficients of the box functions, contain interesting
twistor space structure. For example in $N=4$ gauge theory it has been
shown that the box coefficients of next to MHV amplitudes have planar
support in twistor space, analogous to the behaviour of the tree amplitudes.  
In this paper we have investigated whether
similar behaviour exists for $N<4$  by computing and examining the
box coefficients for all six point $N=1$ amplitudes and certain
classes of $n$ point $N=1$ amplitudes.  It would be interesting
to extend this analysis 
to $N=0$ amplitudes, although in this case, the
box coefficients represent a smaller fraction of the information
contained in the amplitude.
We find that for next to MHV amplitudes these coefficients have
planar support in twistor space, explicitly confirming that the
$N=4$ structure persists to $N=1$.

\vskip 0.6 truecm 
\noindent
{\bf Acknowledgments}

\noindent
This work was supported by a PPARC rolling grant. 
SB would like to thank PPARC for a research studentship.
  
\vfill\eject

\vfill\eject

\end{document}